\documentclass[fleqn,usenatbib,useAMS]{mnras}

\usepackage{graphicx}	% Including figure files
\usepackage{amsmath}	% Advanced maths commands
\usepackage{amssymb}	% Extra maths symbols
\usepackage{multicol}        % Multi-column entries in tables
\usepackage{bm}		% Bold maths symbols, including upright Greek
\usepackage{pdflscape}	% Landscape pages
\usepackage{xcolor}
\usepackage{ulem}
\usepackage{caption}
\usepackage{subcaption}
\defcitealias{Limousin2010}{L10}
\defcitealias{zitrin2015}{Z15}
\defcitealias{Jauzac2015_0416}{J15}
\defcitealias{richard2014}{R14}
\defcitealias{Niemiec2023}{N23}

\usepackage[T1]{fontenc}
\usepackage{ae,aecompl}
\usepackage{newtxtext,newtxmath}
\title[Strong and weak lensing analysis of MACS\,J1423.8$+$2404]{The KALEIDOSCOPE survey : A new strong and weak gravitational lensing view of the massive galaxy cluster MACS\,J1423.8+2404}

\author[N. R. Patel et al. 2024]
{Nency. R. Patel,$^{1,2}$\thanks{Email : nency.r.patel@durham.ac.uk},
Mathilde Jauzac,$^{1,2,3,4}$ 
Anna Niemiec,$^{5}$
David Lagattuta,$^{1,2}$
Guillaume Mahler,$^{1,2,6}$ 
\newauthor
Benjamin Beauchesne,$^{7,8}$
Alastair Edge,$^{1,2}$ 
Harald Ebeling,$^{9}$
and Marceau Limousin$^{10}$
\\
\\
$^{1}$Centre for Extragalactic Astronomy, Department of Physics, Durham University, Durham DH1 3LE, UK \\
$^{2}$Institute for Computational Cosmology, Durham University, South Road, Durham DH1 3LE, UK\\
$^{3}$Astrophysics Research Centre, University of KwaZulu-Natal, Westville Campus, Durban 4041, South Africa \\
$^{4}$School of Mathematics, Statistics \& Computer Science, University of KwaZulu-Natal, Westville Campus, Durban 4041, South Africa \\
$^{5}$LPNHE, CNRS/IN2P3, Sorbonne Université, Université Paris-Cité, Laboratoire de Physique Nucléaire et de Hautes Énergies, 75005 Paris, France \\
$^{6}$ STAR Institute, Quartier Agora - All\'ee du six Ao\^ut, 19c B-4000 Li\`ege, Belgium \\
$^{7}$  Institute of Physics, Laboratory of Astrophysics, Ecole Polytechnique Fédérale de Lausanne (EPFL), Observatoire de Sauverny, CH-1290 Versoix, Switzerland \\
$^{8}$ ESO, Alonso de Cordova 3107, Vitacura, Santiago, Chile \\
$^{9}$Institute for Astronomy, University of Hawaii, 640 N Aohoku Pl, Hilo, HI 96720, USA\\
$^{10}$Aix Marseille Univ, CNRS, CNES, LAM, F-13388 Marseille, France\\
}

\date{Accepted XXX. Received YYY; in original form ZZZ}

\pubyear{2023}

\begin{document}
\label{firstpage}
\pagerange{\pageref{firstpage}--\pageref{lastpage}}
\maketitle

\begin{abstract}
We present a combined strong and weak gravitational-lensing analysis of the massive galaxy cluster MACS\,J1423.8+2404 ($z=0.545$, MACS\,J1423 hereafter), one of the most dynamically relaxed and massive cool-core clusters discovered in the MAssive Cluster Survey at $z>0.5$. We combine high-resolution imaging from the \textit{Hubble Space Telescope} (\textit{HST}) in the F606W, F814W, and F160W pass-bands with spectroscopic observations taken as part of the KALEIDOSCOPE survey with the  Multi-Unit Spectroscopic Explorer mounted on the Very Large Telescope. Our strong lensing analysis of the mass distribution in the cluster core is constrained by four multiple-image systems (17 individual images) within redshift range $1.779<z<2.840$. Our weak-lensing analysis of the cluster outskirts, confined to the \textit{HST} field of view, is based on a background galaxy catalogue with a density of 57\,gal.arcmin$^{-2}$. We measure a projected mass of $M(R<200$\,kpc) = (1.6 $\pm$ 0.05) $\times$ 10$^{14}$\,M$_{\rm\odot}$ from our strong-lensing model, and a projected mass of $M(R<640$\,kpc) = (6.6 $\pm$ 0.6) $\times$ 10$^{14}$\,M$_{\rm\odot}$ when combining with our the weak-lensing constraints. Our analysis of the cluster mass distribution yields no evidence of substructures, confirming the dynamically relaxed state of MACS\,J1423. Our work sets the stage for future analysis of MACS\,J1423 in the upcoming Canadian Near Infrared Imager and Stiltless Spectrograph Unbiased Cluster Survey for the \textit{James Webb Space Telescope}.

\end{abstract}


%%%%%%%%%%%%%%%%%%%%%%%%%%%%%%%%%%%%%%%%%%%%%%%%%%

%%%%%%%%%%%%%%%%% BODY OF PAPER %%%%%%%%%%%%%%%%%%

% The MNRAS class isn't designed to include a table of contents, but for this document one is useful.
% I therefore have to do some kludging to make it work without masses of blank space.
%\begingroup
%\let\clearpage\relax
%\tableofcontents
%\endgroup
%\newpage

\section{INTRODUCTION}
\label{sec:intro}
In the $\Lambda$ Cold Dark Matter ($\Lambda$CDM) cosmological concordance model, where the mass of the Universe is dominated by dark matter, structures form hierarchically. As predicted by the standard model of cosmology \citep{Klypin1983}, the anisotropic collapse of matter on large scales gives rise to large-scale web-like structures known as the cosmic web \citep{1996Natur.380..603B,2012PhRvD..86j3518P,2013MNRAS.430.2200K}. Overall, the cosmic web is a complex distribution consisting of walls, filaments, nodes, and voids. The web itself contains nearly all matter in the Universe, with the majority (both baryonic and dark) residing along the filaments and these nodes.  

Galaxy clusters are the most massive observable objects in our Universe. They form through the accretion of matter from their surroundings and mergers with smaller galaxy groups and clusters. They reside at the intersection of filaments that funnel matter onto them, and are thus called nodes of the cosmic web \citep{2006Natur.440.1137S}. Characterised by extreme masses and dynamical environments, galaxy clusters represent one of the best laboratories to study dark matter and constrain cosmology \citep{2010PhRvD..81d3519K,von_der_linden2014,kelly2014,Applegate2014,Mantz2015,Mantz2016,Jauzac2016,2017MNRAS.467.2913S,hofmann2017,schwinn2018,Jauzac2018,2019MNRAS.488.3646R,grandis2019,shirasaki2020,chiu2023,lyskova2023}. 

So far, dark matter has eluded direct detection, and identifying its nature and key properties remains one of the biggest challenges of modern astrophysics and cosmology. While direct detection is an ongoing objective for particle physicists, astronomers have managed to indirectly observe it through gravitational lensing, i.e., the distortion of the light emitted from a background object by an intervening mass concentration acting as the lens, such as a galaxy cluster. Due to their large masses (the majority of which is dark matter), galaxy clusters strongly deform space-time around them, resulting in deflections of light from distant background sources that manifest by dramatic magnification and distortion of the lensed sources.

Gravitational lensing in galaxy clusters can be mainly observed under two regimes. The strong lensing regime is observed in high density regions of the cluster. In this case, distortions are large, and background galaxies can be observed under the form of gravitational arcs and multiple images. Oppositely, the weak lensing regime is observed in less dense regions and distortions remain very small, necessitating a statistical analysis to extract this lensing signal. Weak lensing generally provides an overall mass estimate of the cluster, and helps trace the distribution of low-mass substructures present in the outskirts of clusters, direct tracers of the past and/or ongoing dynamical events \citep{2011A&ARv..19...47K, Jauzac2016,Jauzac2018}. Using galaxy clusters as gravitational lenses is common practice to study the dynamics and evolution and thus trace their dark matter distribution. This is the main scope of the analysis presented in this paper.

Galaxy clusters can be observed at almost all evolution times. The observer can thus witness different dynamical processes and evolutionary states. In this analysis, we concentrate on dynamically relaxed cluster, i.e. galaxy cluster which tend to show a relatively cool-core, with uniform properties such as temperature and density profiles \citep{2016MNRAS.462..681M,2006ApJ...640..691V}. This cluster population provides insights into the  physics of the intracluster medium (ICM) and its interactions with cluster galaxies \citep{2016MNRAS.456.4020M}. The restriction to clusters with the highest possible degree of dynamical relaxation (for which the assumption of hydro-static equilibrium should be most valid) allows for the most precise test of the CDM model predictions \citep{2007MNRAS.379..209S}. Studying massive relaxed clusters at high redshift ($z>0.5$) are key to understand structure formation and evolution over cosmological timescales as they trace back to around redshift $z=0.5$, i.e. 7 billion years after the Big Bang. One way to do so is to use all-sky surveys to detect all clusters, providing us with the capacity to lead statistical analyses. While the upcoming years are expected to be extremely bright with the successful launch of the \textit{James Webb Space Telescope} on December 25 2021 and the \textit{Euclid Space Telescope} \citep{Racca2016,euclid2022,troja2023} on July 1 2023, the upcoming first light of the Vera Rubin Observatory, and the future launch of the \textit{Nancy Grace Roman Space Telescope}, we here concentrate on observations taken with the largest X-ray survey dedicated to galaxy clusters, the MAssive Cluster Survey \citep[MACS hereafter,][]{2001ApJ...553..668E}.

MACS is an X-ray cluster survey which provides a complete sample of massive and X-ray luminous galaxy clusters at redshifts $z>0.3$. Cluster candidates were selected from the ROSAT (Röntgensatellit) Bright Source Catalogue \citep{1999A&A...349..389V}, based on their X-ray flux and X-ray hardness ratio cuts. The sample included 124 spectroscopically confirmed clusters at $0.3<z<0.7$ thanks to observations taken with the University of Hawaii's 2.2\,m and the Keck 10\,m telescopes. 

In this paper, we concentrate on one of the MACS clusters, MACS\,J1423.8$+$2404 (MACS\,J1423 hereafter). MACS\,J1423 is a massive cluster with a virial mass of $M_{\rm vir}=4.52_{-0.64}^{+0.79}\times 10^{14}$\,M$_\odot$ \citep{2007MNRAS.379..209S} at redshift, $z=0.545$, with an X-ray luminosity $L_{\rm X,bol}=3.7\times 10^{45}$\,erg.s$^{-1}$ \citep{2007ApJ...661L..33E}. MACS\,J1423 is part of the `high-redshift' MACS subsample composed of 12 galaxy clusters at $z>0.5$, and has been the subject of several analyses since its discovery.

\citet{2003ApJ...583..559L} examined it's system's Sunyaev-Zel'dovich (SZ) morphology and galaxy distribution which revealed a cool-core, relaxed cluster. The first gravitational-lensing analysis of MACS\,J1423 was published in \citet{Limousin2010}, followed up by \cite{zitrin2011}, \cite{zitrin2015}, and \citet{merten2015}. All of which confirmed the conclusion of the SZ analysis, MACS\,J1423 is a dynamically relaxed cluster. As noted by \citet[][hereafter L10]{Limousin2010}, MACS\,J1423 is a nearly virialised cluster with very little substructure. The system is a part of the sample targeted by the Cluster Lensing And Supernova survey with \textit{Hubble} \citep[CLASH, Prog.ID: 12790, PI: M. Postman;  ][]{postman2012} and has been used in several statistical studies, where its relaxed dynamical state has been particularly important for cosmological work using the baryon fraction \citep{2003ApJ...583..559L,2004MNRAS.353..457A,2007MNRAS.379..209S}. MACS\,J1423 has also been used to study the triaxiality of galaxy clusters \citep{limousin2013} and was included in several statistical cluster lensing analyses \citep{zitrin2011,limousin2013,Applegate2014,zitrin2015,merten2015}. 

MACS\,J1423 was also observed as part of the KALEIDOSCOPE survey (Prog.ID: 0102.A-0718(A), P.I: A. Edge), the largest snapshot survey of massive clusters, with more than 200 cluster cores observed so far, taken with the Multi-Unit Spectroscopic Explorer (MUSE) spectrograph mounted on the Very Large Telescope (VLT). The target list is comprised of X-ray selected clusters at $z>0.15$ from the ROSAT All Sky Survey (RASS), with the primary sample being the MACS sample. Clusters are selected on the basis of their X-ray luminosities, $L_{\rm X}>3\times 10^{44}$\,erg\,s$^{-1}$ (0.1-2.4\,keV). The field of view of MUSE (1\arcmin$\times$1\arcmin) covers the strong lensing region, i.e. inner core, of most galaxy clusters (typically 15-10\arcsec), and thus provides spectroscopic observations for most galaxies (foreground, cluster or background objects) in the field. KALEIDOSCOPE observations have already proved extremely successful, they have been used for detailed cluster analyses \citep[e.g.,][]{2019MNRAS.483.3082J,mahler2019,2023MNRAS.522.1091L}, including SMACS\,J0723.3-7327, the first released cluster observed by the \textit{JWST} \citep{caminha2022,pascale2022,mahler2023,diego2023}.

We here present a strong and weak lensing analysis of MACS\,J1423 built on the legacy observations of this cluster from space and the 
ground and the use of new mass modelling methods. The paper is organised as follows : Section\,\ref{sec:obs} summarises the data used in this work. The strong and weak lensing analyses are described in Section\,\ref{sec:sl} and Section\,\ref{sec:wl} respectively. Section\,\ref{sec:sl+wl} presents the combined strong and weak lensing analysis, along with the resulting mass distribution of the cluster. Finally, we discuss our results and put them in perspective of previous works on MACS\,J1423 in Section\,\ref{sec:disc}.
Throughout the paper, we assume a flat cosmological model with $\Omega_M$ = 0.3, $\Omega_\Lambda$ = 0.7, and a Hubble constant H$_{0}=70$\,km\,s$^{-1}$Mpc$^{-1}$. In this cosmology, $1\arcsec =6.38$\,kpc at the redshift of the cluster, $z = 0.545$. All magnitudes are given in the AB system \citep{1974ApJS...27...21O}.

\section{OBSERVATIONS}
\label{sec:obs}

\begin{table*}
\centering
    \caption{List of \textit{HST} observations of MACS\,J1423 available. In this work, we use the CLASH observations in the F606W, F814W, and F160W pass-bands. Columns 1 and 2 give the right ascension and declination of the observed field in degrees. Column 3 lists the instruments and filters. Column 4 and 5 give the exposure time in seconds and the observation date, respectively. Finally, columns 6 and 7 provide the programme ID and the PI respectively.}
    \begin{tabular}{c c c c c c c}
    \hline\hline
        R.A.\,(J2000) & Dec.\,(J2000) & Instrument/Filter & Exp. Time (s) & Obs. Date & Prog. ID & PI \\
        \hline\hline     
        215.9524858 & 24.0803087 & ACS/F555W & 4500.000 & 2004-06-16 & 9722 & Ebeling \\
        215.9524858 & 24.0803087 & ACS/F814W & 4590.000 & 2004-06-16 & 9722 & Ebeling  \\
        \hline
        215.9525000 & 24.0803056 & ACS/F814W & 2184.000 & 2006-03-24 & 10493 & Gal-Yam  \\
        \hline        
        215.9495000 & 24.0784722 & WFC3/F225W & 3592.000 & 2013-02-22 & 12790 & Postman \\ 
        
        215.9494593 & 24.0784797 & WFC3/F275W & 3680.000 & 2013-01-13 & 12790 & Postman \\
        
        215.9495000 & 24.0784722 & WFC3/F390W & 1152.000 & 2013-01-13 & 12790 & Postman \\
        215.9494604 & 24.0784607 &  & 1237.000 & 2013-02-22 & 12790 & Postman \\

        215.9495000 & 24.0784722 & ACS/F435W & 1032.000 & 2012-12-31 & 12790 & Postman \\
        215.9496433 & 24.0787853 & & 1066.000 & 2013-03-05 & 12790 & Postman \\

        215.9497915 & 24.0786826 & ACS/F475W & 1092.000 & 2013-02-03 & 12790 & Postman \\
        215.9496433 & 24.0787853 & & 1092.000 & 2013-03-12 & 12790 & Postman \\
        
        215.9497915 & 24.0786826 & ACS/F606W & 1088.000 & 2013-01-13 & 12790 & Postman \\
        215.9495000 & 24.0784722 & & 1032.000 & 2013-02-05 & 12790 & Postman \\

        215.9495000 & 24.0784722 & ACS/F775W & 1032.000 & 2013-01-13 & 12790 & Postman \\
        215.9495000 & 24.0784722 & & 1032.000 & 2013-01-19 & 12790 & Postman \\

        215.9494984 & 24.0785257 & ACS/F850LP & 1065.000 & 2012-12-31 & 12790 & Postman \\
        215.9494702 & 24.0785183 & & 1103.000 & 2013-01-19 & 12790 & Postman \\
        215.9495000 & 24.0784722 & & 1032.000 & 2013-02-03 & 12790 & Postman \\
        215.9495000 & 24.0784722 & & 1032.000 & 2013-03-12 & 12790 & Postman \\

        215.9497638 & 24.0784634 & WFC3/F105W & 1305.869 & 2013-01-22 & 12790 & Postman \\
        215.9495000 & 24.0784722 &  & 1508.802 & 2013-02-05 & 12790 & Postman \\ 

        215.9495000 & 24.0784722 & WFC3/F110W & 1005.802 & 2013-01-19 & 12790 & Postman \\
        215.9494572 & 24.0783802 & & 1005.868 & 2013-03-12 & 12790 & Postman \\
        
        215.9494138 & 24.0784107 & WFC3/F125W & 1005.868 & 2012-12-31 & 12790 & Postman \\ 
        215.9495000 & 24.0784722 & & 1508.802 & 2013-01-01 & 12790 & Postman \\
        
        215.9495000 & 24.0784722 & WFC3/F140W & 1305.869 & 2013-01-22 & 12790 & Postman \\
        215.9494572 & 24.0783802 & & 1005.868 & 2013-02-05 & 12790 & Postman \\

        215.9495377 & 24.0786077 & WFC3/F160W & 1005.868 & 2012-12-31 & 12790 & Postman \\
        215.9497638 & 24.0784634 & & 1005.868 & 2013-01-19 & 12790 & Postman \\
        215.9495000 & 24.0784722 & & 1508.802 & 2013-02-03 & 12790 & Postman \\
        215.9495000 & 24.0784722 & & 1508.802 & 2013-03-12 & 12790 & Postman \\
        \hline\hline 
    \end{tabular}
    \label{tab : hst_obs}
\end{table*}

\begin{table*}
\centering
\caption{Summary of the VLT/MUSE observations of MACS\,J1423 taken as part of the KALEIDOSCOPE Survey. Columns 1 and 2 give the right ascension and declination of the observed field in degrees (J2000). Column 3 gives the redshift of the cluster. Column 4 lists the ESO Programme ID. Column 5 and 6 list the exposure time in seconds and observing date respectively. Column 6 gives the average seeing at the time of observation. The programme ID and PI name are given in column 7 and 8 respectively.}
\begin{tabular}{c c c c c c c c}
\hline\hline
RA (J2000) & DEC (J2000) & z  & Exp. Time (s) & Obs. Date  & Seeing($\arcsec$)  & ESO Programme & PI \\
\hline
215.949458 & 24.078472 & 0.5432  & 2910 & 2019-03-18 & 0.57 & 0102.A-0718(A) & Edge \\
\hline\hline
\end{tabular}
\label{tab:muse_obs}
\end{table*}

\subsection{\textit{Hubble Space Telescope}}

MACS\,J1423 was first observed on 2004 June 16 with the \textit{Advanced Camera for Surveys} (\textit{ACS}) on-board the \textit{Hubble Space Telescope} (\textit{HST}) for 4.5\,ks and 4.6\,ks with the F555W and F814W pass-bands respectively (GO: 9722, PI: Ebeling). 
In 2006, MACS\,J1423 was then imaged again with \textit{ACS} in the F814W pass-band (Prog.ID: 10493, PI : Gal-Yam). 
Finally, MACS\,J1423 is part of CLASH survey \citep{postman2012} and was thus observed again with \textit{ACS} and the \textit{Wide Field camera 3/IR} (\textit{WFC3}) in the F225W, F275W, F390W, F435W, F475W, F606W, F775W, F850LP, F105W, F110W, F125W, F140W and F160W pass-bands. A list of all available observations of MACS\,J1423 with \textit{HST} is given in Table\,\ref{tab : hst_obs}.

For the analysis presented in this paper, we use the CLASH data products in the F606W, F814W and F160W pass-bands\footnote{\url{https://archive.stsci.edu/missions/hlsp/clash/macs1423/data/hst/scale_30mas/}}. For a galaxy cluster at redshift $z=0.545$, it is important to use filters that capture the rest-frame optical light, particularly the 4000 $\mathring{A}$ break to identify the old and red galaxies in the cluster. Given the cluster redshift, the combination of F606W and F814W is best for getting this old population (see Sect.\,\ref{sec:cluster_members}). Additionally, F160W, in combination with F606W and F814W, is well-suited for identifying background galaxies in the cluster (see Sect.\,\ref{sec:bg}). The mosaics were produced using the \textsc{MosaicDrizzle} pipeline \citep{2002hstd.book.....K,2011ApJS..197...36K}. These consist of drizzled science images in counts/s, along with associated inverse variance weight images, which are scaled so that they can be used to create the rms mosaics that are in the same units as the drizzled science image mosaics. 

\subsection{\textit{Chandra} X-ray Observatory}
\label{sec:chandra}
MACS\,J1423 was first observed in the X-rays on 2001 June 1 with the \textit{Advanced CCD Imaging Spectrometer (ACIS)} onboard the \textit{Chandra X-ray Observatory} for 18.53\,ks (Obs.ID: 1657, P.I : Vanspey-broeck). The cluster was observed again on 2003 August 18 for 115.57\,ks (Obs.ID: 4195, P.I: Allen). We reduce the data as in \citet{Beauchesne2024} with the \textit{Chandra} pipeline \textsc{ciao}\footnote{\url{https://cxc.cfa.harvard.edu/ciao/}} $4.15$ \citep{ciao2006} and \textsc{caldb} $4.10.7$ and produce a counts map in the broad energy band (i.e. [0.5,7] keV). These X-ray observations are used for comparison with our lensing analysis.

\subsection{Ground-based Spectroscopy and Photometry}
\label{sec:spec_photo_z}

\subsubsection{Subaru \& Canada France Hawaii Telescope (CFHT)}
MACS\,J1423 was observed in the B, V, $R_c$, $I_c$ and z' bands with the SuprimeCam wide-field imager on the Subaru telescope \citep{Miyazaki2002}, and in the u$^{*}$ and K bands with MegaCam and Wide-field Infrared Camera (WIRCam) on the Canada France Hawaii Telescope (CFHT). The data reduction techniques were adapted to deal with the special characteristics of the Suprime-Cam and MegaCam; details are given by \citet{Donovan2007}. The resulting imaging data are used to compute photometric redshifts for all galaxies in a 0.5 $\times$ 0.5 deg$^{2}$ field following the methodology described in \citep{Ma2008}. To estimate the spectral energy distribution (SED) for all objects within the field of view and to ensure that the images for all bands have the same effective spatial resolution of 1$\arcsec$, the imaging data in various pass-bands are seeing-matched using the technique described in \citet{Kartaltepe2008}. 
\textsc{Source-Extractor} \citep[][or \textsc{SExtractor}]{1996A&AS..117..393B} was used in dual image mode to create the object catalogue with R-band image as the reference detection image. 
An adaptive SED fitting code Le Phare \citep{Arnouts1999,Ilbert2006,Ilbert2009} was used to determine the photometric redshifts for galaxies with $m_{\rm R_c}$ < 24.0. To reduce the fraction of catastrophic errors and to mitigate the systematic trends in the difference between spectroscopic and photometric redshifts, Le Phare adjusts the photometric zero points by using galaxies with spectroscopic redshifts as training points. More details are given in \citet{Ma2008}.

\subsubsection{W.M. Keck Observatory}
\citetalias{Limousin2010} observations were taken with the Low Resolution Imager and Spectrograph \citep[LRIS,][Prog.ID: H26aL, PI:Henry]{1995PASP..107..375O} and the Deep Imaging Multi-Object Spectrograph \citep[DEIMOS, ]{1997SPIE.2871.1107C} on the 10\,m Keck Telescope on Mauna Kea.

LRIS observations provided spectroscopic redshifts for 3 multiple image systems in the cluster. The reduction of these data is described in detail in \cite{2002ApJ...576..720K}. Redshifts of the galaxies in the cluster field were obtained from LRIS and DEIMOS and published by \citetalias{Limousin2010}. Cluster members were defined as galaxies with  redshifts within $\pm$0.05 of the cluster redshift $z_{\rm cluster}=0.545$.

\subsubsection{The Very Large Telescope}
\label{sec:muse}

Panoramic, integral-field spectroscopic coverage of MACS\,J1423 was obtained with MUSE \citep{2010SPIE.7735E..08B} at the VLT, as part of the KALEIDOSCOPE cluster survey (P.ID: 0102.A-0718(A), PI: A. Edge). Observations were taken on 2019 March 17 during grey time, under clear sky conditions, with an average seeing of 0.57\arcsec and three 970\,s exposures, spanning the wavelength range 4800-9300$\mathring{A}$. To reduce systematic effects of bad pixels and variable detector sensitivity, a small (0.5\arcsec) dither was applied between exposure. Each exposure was also rotated by 90 degrees from the previous one. However, to maximize integrated depth, the centres of each exposure were coincident -- save for the small dither offset. Thus, the total exposure time over most of the field of view (1$\times$1\arcmin$^{2}$) is 2910\,s. Specific details of these observations are given in Table\,\ref{tab:muse_obs}.

The data reduction and spectroscopic extraction were performed by our team following the method presented in \cite{lagattuta2022}, which we briefly summarize here. The reduction occurs in two part. First, we follow the procedure described in the public MUSE Data Reduction Pipeline User Manual\footnote{\url{https://www.eso.org/sci/software/pipelines/muse/}}. Namely, we apply bias and flat-field correction, wavelength calibration, line-spread function (LSF) modelling, and sky-subtraction to 2-dimensional ``pixel table'' configurations of each exposure, using nightly calibration files that accompany the raw science frames. During this step, we also calibrate the flux levels of the data and correct for telluric absorption using a standard star observed immediately after the science frames. Following these corrections, we interpolate the pixels of each exposure onto a regular grid, combine all exposures together, and restructure the data into a cube. After the initial reduction phase we apply two additional reduction calibrations to the combined cube directly. We model and correct low-level flat-field residuals caused by intra-cluster light (known as ``auto-calibration'', see e.g. section 2.5 in \citealt{richard21}). Additionally, we apply the Zurich Atmospheric Purge algorithm (ZAP; \citealt{soto2016}) to remove any significant sky-line residuals that remain following sky subtraction.

After reducing the cube, we select targets for spectroscopic extraction using two distinct but complementary techniques. In the first method we identify objects in the \textit{HST} images that fall in the MUSE Field of View (FoV). We measure the position and shape of these objects using \textsc{SExtractor}, then convolve the shapes with the MUSE PSF to better match their size in the data cube. The second method also relies on \textsc{SExtractor}, but this time it is run on the cube directly, to identify prominent emission-line objects that may not have significant stellar continuum. We cross-match the two samples to remove any duplicate detections and extract a spectrum for all remaining targets using the optimally-weighted \citet{horne1986} procedure. Following extraction we run each spectrum through the redshift-fitting software \textsc{marz} \citep{Hinton2016}, which provides an initial redshift guess for each object. We then visually inspect the results, accepting or modifying the best-fit \textsc{marz} redshift as needed.

The final redshift catalogue for MACS\,J1423 contains 66 entries, consisting of two stars, three foreground galaxies ($z < z_{\rm cluster}$), 39 cluster members and 16 background galaxies ($z > z_{\rm cluster}$). A list of cluster member galaxies is given in Table\,\ref{tab:muse}, while multiple-image systems are summarised in Table\,\ref{tab:mulimg}. We note that these results highlight MUSE's remarkable efficiency as a spectroscopic detector in crowded (i.e., cluster-core) fields, allowing us to cleanly measure the redshifts of not only multiply-imaged galaxies, but also intervening objects along the line-of-sight.

\begin{table}
\centering
\caption{Cluster members spectroscopically confirmed by VLT/MUSE observations in MACS\,J1423. Columns 1 and 2 give the right ascension and declination in degrees (J2000). Column 3 gives their respective measured spectroscopic redshift with VLT/MUSE.}
\begin{tabular}{c c c}
\hline\hline
RA (J2000)         & DEC (J2000)         & z      \\
\hline\hline
215.9494838 & 24.0784459 & 0.545 (BCG) \\
215.9523294 & 24.0707291 & 0.539 \\
215.9427198 & 24.0727851 & 0.537 \\
215.9539395 & 24.0714300 & 0.539 \\
215.9586154 & 24.0710327 & 0.533 \\
215.9477283 & 24.0727531 & 0.546 \\
215.9525992 & 24.0742691 & 0.541 \\
215.9517541 & 24.0741379 & 0.554 \\
215.9489410 & 24.0763359 & 0.537 \\
215.9462858 & 24.0754164 & 0.562 \\
215.9474872 & 24.0761680 & 0.547 \\
215.9493174 & 24.0769683 & 0.555 \\
215.9493511 & 24.0767400 & 0.560 \\
215.9448420 & 24.0770788 & 0.545 \\
215.9497831 & 24.0779724 & 0.523 \\
215.9496876 & 24.0773563 & 0.548 \\
215.9484518 & 24.0770044 & 0.517 \\
215.9453596 & 24.0784154 & 0.546 \\
215.9565067 & 24.0779951 & 0.530 \\
215.9516898 & 24.0782917 & 0.537 \\
215.9581679 & 24.0787077 & 0.552 \\
215.9584832 & 24.0793582 & 0.535 \\
215.9448590 & 24.0785557 & 0.533 \\
215.9462005 & 24.0795718 & 0.531 \\
215.9433649 & 24.0802384 & 0.543 \\
215.9510899 & 24.0808902 & 0.532 \\
215.9582277 & 24.0812035 & 0.552 \\
215.9507082 & 24.0813282 & 0.551 \\
215.9403157 & 24.0823716 & 0.560 \\
215.9502609 & 24.0842187 & 0.550 \\
215.9462506 & 24.0848389 & 0.546 \\
215.9472276 & 24.0841258 & 0.549 \\
215.9537427 & 24.0845512 & 0.554 \\
215.9494303 & 24.0852384 & 0.552 \\
215.9451689 & 24.0847458 & 0.545 \\
215.9474839 & 24.0856535 & 0.546 \\
215.9499363 & 24.0865008 & 0.542 \\
215.9488118 & 24.0860833 & 0.563 \\
215.9532771 & 24.0868116 & 0.540 \\
\hline\hline \\
\end{tabular}
\label{tab:muse}
\end{table}

\section{STRONG GRAVITATIONAL LENSING ANALYSIS}
\label{sec:sl}
\subsection{Multiple image systems}
\label{sec:mul_img}
The position, shape and geometry of multiple images of background galaxies will depend on the mass distribution in the galaxy cluster lens, the distance of the source galaxies, together with the alignment between the observer, the cluster and the lensed sources. As a result, in order to constrain the mass distribution of MACS\,J1423, we use the positions of the identified multiple image systems as constraints for our strong lensing mass model, i.e. in the core of the cluster.

\subsubsection{Previous Work}
\label{sec:previous mult_img}
Prior to this analysis, three multiple image systems were reported in \citetalias{Limousin2010}, and another one in \citet[][hereafter, Z15]{zitrin2015}. These systems are listed in Table\,\ref{tab:mulimg} as Systems 1-\,4. For \citetalias{Limousin2010}, systems 1 and 2 belong to a single background galaxy, resolved in two components with multiple images organised in an Einstein-cross configuration. A central fifth image is also predicted at the position of the brightest cluster galaxy (BCG) located at the very centre of MACS\,J1423. Due to its large halo of stars, and the fact that central images are usually de-magnified, multiple images located in this region are generally very difficult, if not impossible, to detect. On the other hand, system 3 is seen in a naked-cusp configuration, with three images on the same side of the cluster. Another multiple image system is found by \citetalias{zitrin2015}, composed of five images, and which includes a radial arc. System 1 and 2 in \citetalias{Limousin2010} are taken as a single system in \citetalias{zitrin2015} without the central radial arc (image 1.5 in \citetalias{Limousin2010}). All systems can be seen in Fig.\,\ref{fig:mul_img}.

\begin{table}
\caption{Multiple image systems used as constraints in our strong-lensing analysis of MACS\,J1423. Column 1 lists the ID of the multiple images. Columns 2 and 3 give the right ascension and declination in degrees (J2000). Column 4 gives the redshift of each system. Redshifts without error bars correspond to spectroscopic measurements while the presence of error bars highlight the redshifts optimised by our mass model.  
While spectroscopic redshifts were already measured with LRIS and DEIMOS by \citetalias{Limousin2010}, we highlight with an $\ast$ the systems for which we confirm spectroscopic redshifts with MUSE observations. Column 5 gives the magnification estimated from our strong lensing mass model for each multiple images.}
\centering
\begin{tabular}{c c c c c}
\hline\hline
ID  & RA (deg)          & DEC (deg)       & $z$   & $\mu$  \\
\hline\hline
1.1 & 215.9577155 & 24.0746537 & 2.84  & 3.49 $\pm$ 0.52  \\
1.2 & 215.9494609 & 24.0818270 & 2.84 & 7.08 $\pm$ 1.45 \\
1.3 & 215.9444674 & 24.0806542 & 2.84 & 9.25 $\pm$ 2.04 \\
1.4 & 215.9459355 & 24.0763898 & 2.84 & 10.89 $\pm$ 1.79 \\
1.5 & 215.9497029 & 24.0791219 & 2.84  & 0.99 $\pm$ 0.36 \\
\hline
2.1 & 215.9576456 & 24.0745519 & 2.84 & 3.50 $\pm$ 0.52  \\
2.2 & 215.9495433 & 24.0817288 & 2.84 & 7.15 $\pm$ 1.49 \\
2.3 & 215.9443426 & 24.0803974 & 2.84 & 10.97 $\pm$ 2.39 \\
2.4 & 215.9458112 & 24.0763495 & 2.84 & 11.87 $\pm$ 1.96  \\
\hline
3.1\textsuperscript{*} & 215.9465404 & 24.0836281 & 1.779 & 8.82 $\pm$ 1.53 \\
3.2\textsuperscript{*} & 215.9523020 & 24.0831945 & 1.779 & 16.74 $\pm$ 3.82 \\
3.3\textsuperscript{*} & 215.9551508 & 24.0811756 & 1.779 & 13.71 $\pm$ 2.45 \\
\hline
4.1 & 215.9481287 & 24.0771353 & $1.797\pm0.062$ & 5.07 $\pm$ 1.09 \\
4.2 & 215.9558429 & 24.0771761 & $1.797\pm0.062$ & 4.15 $\pm$ 0.52 \\
4.3 & 215.9502463 & 24.0821356 & $1.797\pm0.062$ & 11.79 $\pm$ 3.18 \\
4.4 & 215.9459358 & 24.0817256 & $1.797\pm0.062$ & 5.57 $\pm$ 0.73  \\
4.5 & 215.9490200 & 24.0780830 & $1.797\pm0.062$ & 1.48 $\pm$ 0.38 \\
\hline\hline
\end{tabular}
\label{tab:mulimg}
\end{table}

\subsubsection{This work} 
\begin{figure*}
    \centering
    \includegraphics[width=\linewidth]{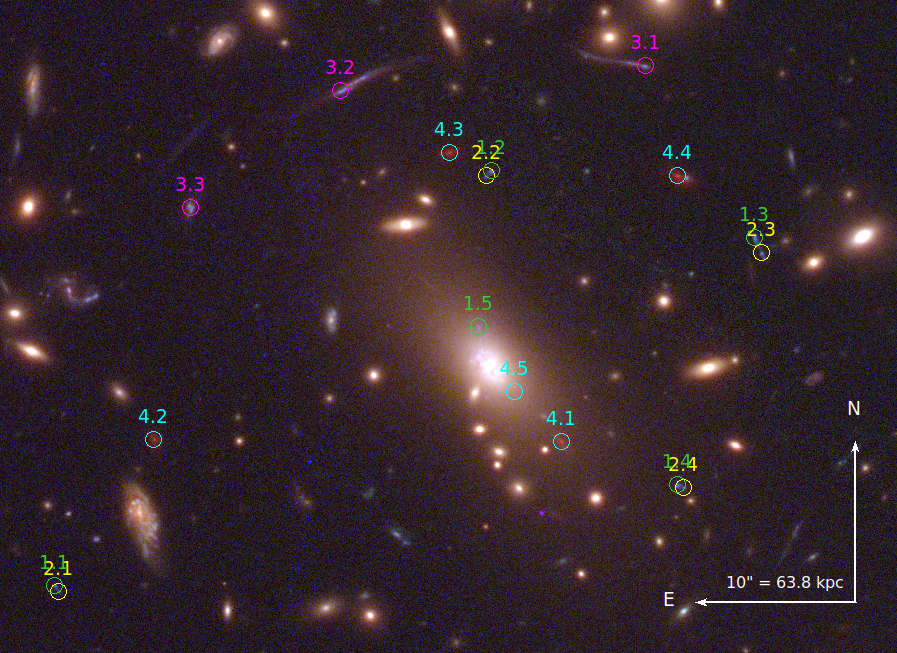}
    \caption{\textit{HST} colour composite image of the central region of MACS\,J1423 created using the F606W for blue, F814W for green and F160W for red colours. Multiple image systems are highlighted with coloured circles and ID following the list given in Table\,\ref{tab:mulimg}. The five multiple images of system 1 are shown in green, the four multiple images of system 2 in yellow, the three multiple images of system 3 in magenta, and the five multiple images of system 4 in cyan.}
    \label{fig:mul_img}
\end{figure*}

The four systems discussed in the previous subsection are used. We combine the identifications from the two analyses of \citetalias{Limousin2010} and \citetalias{zitrin2015} (see Table ~\ref{tab:mulimg} and Fig.\,\ref{fig:mul_img}), and thus use the 4 systems discussed in Sect.\,\ref{sec:previous mult_img}.

In contrast to \citetalias{Limousin2010}, we add the fourth system of \citetalias{zitrin2015}, system 4 in Table~\ref{tab:mulimg}, while systems 1, 2 and 3 are considered in the same manner in \citetalias{Limousin2010}. Adding to that, systems 1 and 2 correspond to the same multiply imaged galaxy, but for which \citetalias{Limousin2010} identified 2 components, thus we split it in two sets of constraints. We then use the central radial arc of system 1 as done by \citetalias{Limousin2010}.

To summarise, we use four multiple image systems: system 1 with five multiple images, including the central radial arc (image 1.5), system 2 with four multiple images, system 3 with three multiple images, and system 4 with 5 multiple images, including another central radial arc. 
System 1, 2 and 3 are all spectroscopically confirmed. However, the MUSE observations are not of enough quality for us to measure a spectroscopic redshift with high confidence for system 4. We obtained a low confidence measurement of $z=1.878$ which is discussed in more detail in Sect.\,\ref{sec:para_model}, however we let the redshift of this system free to vary in our mass model.

\subsection{Cluster Members}
The mass modelling technique used in this paper account for the matter contribution from cluster members. Consequently, a well defined catalogue of cluster members is needed. We here present the process of getting such a catalogue.

\label{sec:cluster_members}
\subsubsection{Previous Work}

\citetalias{Limousin2010} obtained spectroscopic redshift measurements with LRIS and DEIMOS as described in Sect.\,\ref{sec:obs}. These observations provided spectroscopic confirmation of multiple image systems, together with spectroscopy of cluster galaxies within the field of MACS\,J1423, and allowed them to build a cluster member catalogue. Galaxies with redshifts within $\pm$\,0.05 of the cluster redshift, $z_{\rm cluster}=0.545$, were considered cluster members. The final cluster member catalogue was composed of 30 galaxies, including the Brightest Cluster Galaxy (BCG)

\begin{figure}
\includegraphics[width=\linewidth]{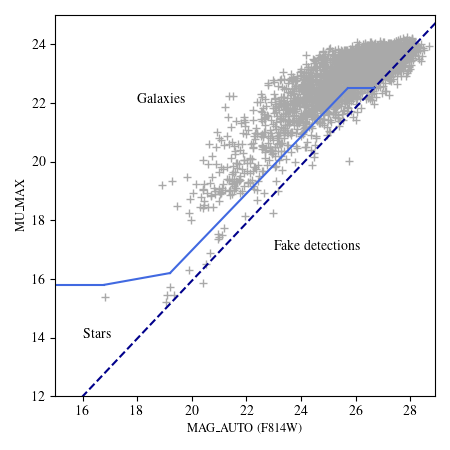}
\caption{Magnitude (MAG$\_$AUTO) vs surface brightness (MU$\_$MAX) diagram for MACS\,J1423 in the F814W pass-band. Such a diagram allows us to disentangle stars and fake detections from galaxies. The blue polygon highlights the sequence of stars and the limit for fake detections. The grey markers within the polygon are the star sequence and the ones outside are objects identified as galaxies in the cluster.}
\label{fig:stars}
\end{figure}

\begin{figure}
    %\hspace*{-1.cm}    
    \begin{subfigure}{\textwidth}
        \includegraphics[width=0.5\textwidth]{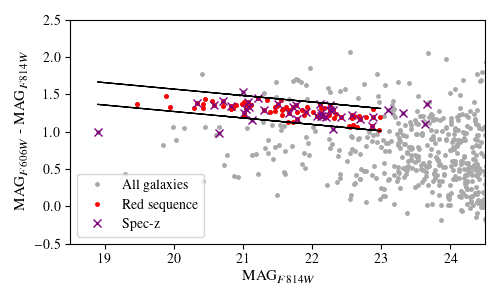}    
    \end{subfigure}\hfill
    \begin{subfigure}{\textwidth}
        \includegraphics[width=0.5\textwidth]{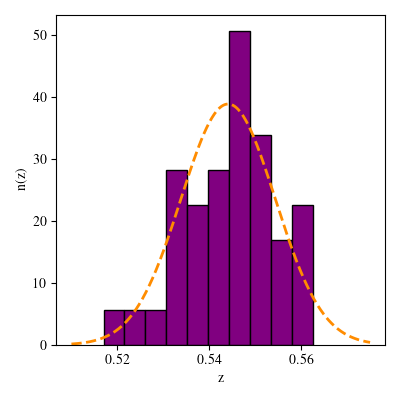}
    \end{subfigure}
    \caption{Top: Colour (mag$_{\rm F606W}$ - mag$_{\rm F814W}$) vs magnitude (m$_{\rm F814W}$) diagram for MACS\,J1423. All sources from the source extraction done in this work are shown in grey, and the red sequence galaxies are highlighted in red marked within the black lines. The cluster members with spectroscopic redshifts are marked by purple cross. We consider all galaxies within the 2\,$\sigma$ of the linear fit as cluster members. Bottom : Histogram of the redshift of spectroscopically confirmed cluster members within MUSE data. The best Gaussian fit is overlaid in orange dashed line.}
    \label{fig:cluster_members}
\end{figure}

\subsubsection{This work}
Galaxy clusters are known to have a well defined and very regular population of early type elliptical and lenticular galaxies, called the red sequence \citep{Gladders2000}. These red sequence galaxies show a tight relation between their colour and magnitude \citep{Baum1959,Visvanathan1977}. \textsc{SourceExtractor} \citep[or \textsc{SExtractor}, ][]{1996A&AS..117..393B} is used in dual mode to extract the F606W and F814W photometry for cluster galaxies. The sequence of stars is identified using the surface brightness and magnitude of the detected sources (the MAG$\_$AUTO vs MU$\_$MAX diagram) as shown in Fig.\,\ref{fig:stars}, and then removed from the catalogue. A colour-magnitude diagram is then created to identify the red sequence itself as shown in Fig.\,\ref{fig:cluster_members}. Sources considered as bad detections (with a quoted magnitude of 99 from \textsc{SExtractor}) are removed along with sources at the edges of the \textit{HST}/ACS field of view. A 2$\sigma$ clipping fit is used to fit the red sequence, and a visual inspection is performed to remove any remaining contamination such as merging non-cluster galaxies in the sample. Any remaining artefacts such as star spikes and edge effects are discarded from our catalogue. Cluster members in the inner region of the cluster (\textit{R}$<$260\,kpc) are confirmed using spectroscopic redshifts measured with our VLT/MUSE observations (see Sect.\,\ref{sec:muse}).
Our final cluster member catalogue contains 66 galaxies, out of which 39 are spectroscopically confirmed, including the Brightest Cluster Galaxy (BCG).

\subsection{Parametric mass modelling}
\label{sec:para_model}
In the cluster core, where we assume light traces, i.e. that dark matter halos (at both cluster and galaxy-scales) follow the light distribution, we use a parametric mass model. The total mass distribution in this region is decomposed into large and small scale components to model cluster and galaxy-scale matter distributions respectively. Each component is modelled assuming a dual Pseudo Isothermal Elliptical density profile \citep[dPIE,][]{2007arXiv0710.5636E} 
characterised by seven parameters: position ($\alpha$, $\delta$), position angle ($\theta$), ellipticity (e), velocity dispersion ($\sigma$), and two scale radii, the core and the cut radii (\textit{r}$_{\rm core}$ and \textit{r}$_{\rm cut}$). The 3D density distribution of the dPIE is written as 
\begin{equation}
    \rho(r) = \frac{\rho_{0}}{(1 + r^{2}/r^{2}_{\rm{core}})(1 + r^{2}/r^{2}_{\rm{cut}})}, \,\,\, \rm{with}\, \textit{r}_{\rm{cut}} > \textit{r}_{\rm{core}}.
\label{eq:dpie}
\end{equation}
where $\rho_0$ is the central density of the core. To reduce the number of free parameters in the lens model, position, ellipticity and angle position of galaxy-scale components are fixed to follow their light distribution. In addition, Faber-Jackson empirical scaling relations \citep{Faber1976} are used for the remaining parameters. In this case, we consider that for each cluster galaxy, the core radius, \textit{r}$_{\rm core}$, the cut radius, \textit{r}$_{\rm cut}$ and the velocity dispersion, $\sigma_0$, scale with the galaxy luminosity, L, as follows:
\begin{equation}
\label{eq:scaling_relation}
\begin{split}  
\sigma_{0} & = \sigma_{0}^{\star} \left(\frac{L}{L_{\star}}\right)^{1/4}, \\
r_{\rm core} & = r_{\rm core}^{\star} \left(\frac{L}{L_{\star}}\right)^{1/2}, \\
r_{\rm cut} & = r_{\rm cut}^{\star} \left(\frac{L}{L_{\star}}\right)^{\alpha}
\end{split}
\end{equation}
where L\textsuperscript{$\star$} is the typical luminosity of an elliptical galaxy at the cluster redshift, and \textit{r}$_{\rm core}^{\star}$, \textit{r}$_{\rm cut}^{\star}$, $\sigma_0^{\star}$ are the dPIE parameters. $\alpha$ is the slope which is fixed to 1/2, following the work by \cite{Niemiec2020,jullo2009}.

The publicly available lens modelling software \textsc{Lenstool} is used for the mass reconstruction. There is one cluster scale halo in our model, for which all the parameters are let free to vary except for the cut radius, which generally lies outside the strong lensing region making it difficult to be constrained by our model and is thus fixed to \textit{r}$_{\rm cut}$\,=\,1000\,kpc. For the small-scale halos, we have 65 cluster members, modelled using scaling relations with $mag_0 = 20.05$ at the MACS\,1423 cluster redshift in the ACS/F814W passband. Their positions are fixed to that of cluster galaxies, along with their position angle and ellipticity. However, core radii are fixed to a very small value of \textit{r}$_{\rm core}^{\star}$\,=\,0.15\,kpc, while cut radii and velocity dispersions are let free to vary with 5 < \textit{r}$_{\rm cut}^{\star} < 50$\,kpc and 100 < $\sigma_{0}^{\star}$ < 250 km\,s$^{-1}$ respectively. The BCG in the cluster is modelled separately from the rest of cluster members, since extremely luminous central cluster galaxies often not follow the scaling relations given in equation\,\ref{eq:scaling_relation} \citep{2013ApJ...765...24N,2013ApJ...765...25N}. 

\begin{figure*}
    \centering
    \includegraphics[width=\linewidth]{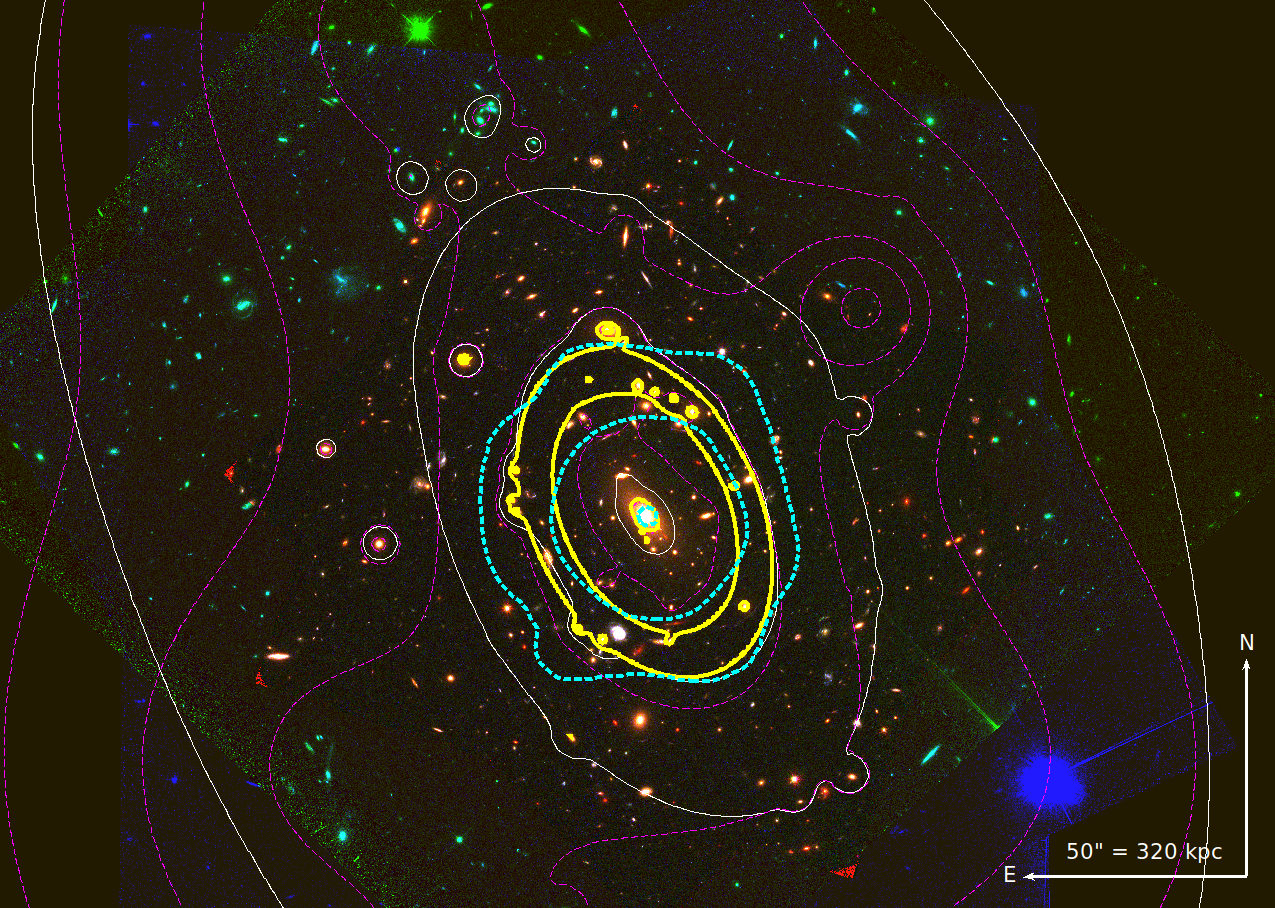}
    \caption{\textit{HST} colour composite image of the central region of MACS\,J1423 created using the F606W for blue, F814W for green and F160W for red colours. The thick solid yellow and thin dashed magenta contours show the mass distribution obtained with our strong-lensing and weak-lensing mass reconstruction as described in Sect.\,\ref{sec:para_model} and Sect.\,\ref{sec:wl_modeling} respectively. The mass contours obtained with our combined mass model, as presented in Sect.\,\ref{sec:sl+wl}, are shown with thin solid white contours. The thick dashed cyan contours represent the X-ray surface brightness from the \textit{Chandra} observations.}
    \label{fig:pot_contours}
\end{figure*}

The free parameters of the different potentials are optimised, using the positions of strongly lensed multiple images as constraints. The goodness of our model to meet the observational constraints is measured with root mean square or rms and $\chi^{2}$. The rms is the difference between observed multiple image positions and predicted position from the model. For the model described, consisting of one large scale halo and 66 small-scale halos, constrained by the 17 multiple images discussed in Sect.\,\ref{sec:mul_img}, we obtain an rms for our best-fit mass model of 0.90\arcsec, and a reduced $\chi_{\nu}^{2}$ of 1.13. The model predicts a redshift of $z=(1.79\pm 0.06)$ for system 4, which agrees well with the tentative spectroscopic redshift derived from the VLT/MUSE observations ($z=1.878$, see Sect.\,\ref{sec:muse}). The best-fit parameters of our strong lensing mass model are listed in Table\,\ref{tab:sl}. The mass contours for this strong-lensing only model is shown in yellow colour in figure\,\ref{fig:pot_contours}. The corresponding surface mass density of MACS\,J1423 as a function of distance from the cluster center is shown in Fig.\,\ref{fig:sdp1} in magenta colour. We measure a projected mass enclosed in the strong lensing region of \textit{M}(${R}<200\,{\rm kpc})=(1.83\pm 0.04)\times$10$^{14}$\,M$_{\rm\odot}$.

\begin{table*}
\caption{Best fit parameters for our strong lensing mass model of the core of MACS\,J1423. Values given in brackets are fixed in our model, while the ones quoted with error bars are being optimised. The coordinates of the cluster and BCG halos are given in arcseconds relative to the cluster center (R.A., Dec)$=$(215.94944 24.078499). The log-likelihood (ln($\mathcal{L}$)), $\chi^{2}$, $\chi_{\nu}^{2}$ and rms are given in the bottom of the table.}
\centering
    \begin{tabular}{c c c c c c c c}
    \hline\hline 
    Components & $\Delta$R.A.(\arcsec) & $\Delta$Dec (\arcsec) & \textit{e} & $\theta$(deg) & $r_{\rm core}$ (kpc) & $r_{\rm cut}$ (kpc) & $\sigma_0$ (km\,s$^{-1}$)
    \\
    \hline
    Cluster Halo & -1.2 $\pm$ 0.6 & -0.5 $\pm$ 1.1 & 0.4 $\pm$ 0.04 & 117 $\pm$ 1 & 137 $\pm$ 23 & [1000] & 1186 $\pm$ 57 \\
    \\
    BCG & [-0.07] & [-0.22] & [0.62] & [310] & [0.15] & 49 $\pm$ 40 & 442 $\pm$ 30  \\
    \\
    L\textsuperscript{*} Galaxy & - & - & - & - & [0.15] & 22 $\pm$ 11 & 211 $\pm$ 30 \\
    \hline
    ln($\mathcal{L}$) = -36.17 & $\chi^{2}$ = 17.03 & $\chi_{\nu}^{2}$ = 1.13 & rms = 0.90\arcsec \\ 
    \hline\hline
    \end{tabular}
\label{tab:sl}
\end{table*}

\section{WEAK GRAVITATIONAL LENSING ANALYSIS}
\label{sec:wl}

\subsection{Weak lensing catalogue}
In the weak lensing regime, the shapes of lensed background sources carry the lensing signal and as a result are used to constrain the mass distribution. Such signal is weak and thus difficult to measure for a single background source. Therefore, a statistical analysis is necessary to extract this faint signal. We here present the construction of the weak lensing background galaxy catalogue for MACS\,J1423. The source detection, photometry and shape measurements are performed with the stacked images in the F606W, F814W, and F160W pass-bands. We refer the reader to \cite[][J12 and J15 respectively hereafter]{Jauzac2012,2015MNRAS.452.1437J} for more details on the method used to generate the weak lensing catalogue, and only give here a summary of the different steps.

\subsubsection{The ACS source catalogue}
\label{sec:acs_cat}
To detect sources and measure their shapes, we use the \textit{HST}/ACS F814W filter. The publicly available weak lensing shape measurement algorithm, \textsc{pyRRG},\footnote{https://pypi.org/project/pyRRG/} based on \cite{2000ApJ...536...79R} and presented in \cite{2019arXiv191106333H},  is used for the construction of the catalogue, including the measurements of shapes.
The first step consists in using the `hot-cold' method in \citet{2007ApJS..172..219L} to optimally extract sources from the image using \textsc{Source-Extractor} \citep{1996A&AS..117..393B}. 
In this method, two \textsc{SExtractor} scans of the image are performed: the hot scan to find the smaller and fainter sources, and the cold scan to find the larger and brighter sources using different minimum. The two catalogues are then combined. 
Following the detection of `hot-cold' sources, the sample is cleaned by removing duplicate detections and any sources close to stars or saturated pixels. The 'hot-cold' method ensures we keep the larger objects. Based on the distribution of sources in the magnitude (MAG$\_$AUTO) vs peak surface brightness (MU$\_$MAX) plane, sources are classified as stars, galaxies and fake detections \citep[see ][\citetalias{Jauzac2015_0416,richard2014} for more details]{2007ApJS..172..219L}. 
Fake detections are then removed from our catalogue, and the sequence of stars is kept for the modelling of the point spread function (PSF), mandatory step in the measurement of shapes.

\subsubsection{Shape measurements}
\label{sec:shapes}
The next step consists on measuring shapes of background objects. This necessitates accounting for instrumental effects, and the position of the telescope, which will impact the true shape of galaxies. \textsc{pyRRG} was constructed accordingly, and is thus correcting for such effects. It models the PSF of individual exposures for all the images considered, and corrects all galaxy shapes for it. It measures the PSF from the stars in each individual exposures, comparing the second and fourth order moments to various \textsc{Tiny Tim} models of the PSF \citep{2011SPIE.8127E..0JK}. \textsc{Tiny Tim} is an easy-to-use standard \textit{HST} modelling software which, thanks to detailed information on the telescope and instruments, models the PSF of the different instruments. It provides access to PSF models that properly matches the conditions of \textit{HST} observations under different conditions. 
\textsc{pyRRG} combines the best-fit PSF over all the individual exposures to get a final PSF model that matches the observations at hands. Having the modelled PSF, the next step is to then correct galaxy moments from it, and measure the shape of these galaxies.

In \textsc{pyRRG}, the ellipticity, $\mathrm{e = (e_1,e_2)}$, with $e_{1}$ and $e_{2}$ the two components of ellipticity, and the size parameter, \textit{d}, are defined as
\begin{equation}
    \begin{split}
        e_1 = \frac{I_{xx} - I_{yy}}{I_{xx} + I_{yy}}, \\
        \\
        e_2 = \frac{I_{xy}}{I_{xx} + I_{yy}}, \\ 
        \\
        d  = \sqrt{\frac{I_{xx} + I_{yy}}{2}}
    \end{split} 
\end{equation}
where \textit{I$_{ij}$} are the second order weighted Gaussian moments. The shear estimator, $\tilde{\gamma}$, is measured using the ellipticity together with the polarisability, \textit{G}, 
\begin{equation}
    \tilde{\gamma} = C\frac{e}{G}
\end{equation}
where \textit{C} is the calibration constant. The polarisability, \textit{G}, is calculated using equation\,28 in \citet{2000ApJ...536...79R}, and the calibration constant, \textit{C}, is fixed to 0.86, as given in \citet{2007ApJS..172..219L}.

\subsubsection{Background galaxy selection}
\label{sec:bg}
We now have a weak lensing catalogue with all shapes of galaxies measured with \textsc{pyRRG}. However, this catalogue is still contaminated by foreground and faint cluster galaxies.
The weak lensing signal is carried by background galaxies, but can be diluted by the presence of cluster members and/or foreground sources. Despite removing cluster member galaxies from the catalogue, the catalogue remains contaminated by faint, small and bluer cluster galaxies (mimicking the colour of background objects), and foreground objects for which redshifts measurements are not available. We thus use a colour-colour selection for galaxies without photometric and/or spectroscopic information as done by \citetalias{Jauzac2015_0416}. In the case of MACS\,J1423, we place our selection in the ($\rm{mag_{F435W}-mag_{F814W}}$) \textit{vs} ($\rm {mag_{F435W}-mag_{F606W}}$) colour-colour parameter space. In the colour-colour diagram, regions dominated by the unlensed population (cluster and foreground galaxies) are identified using the available spectroscopic and/or photometric redshifts. The weakly lensed galaxies are then separated from the unlensed population by a defined polynomial region using this calibrated selection (see \citetalias{Jauzac2015_0416} for more details).
As a final step, to remove galaxies with ill-determined shape parameters. The final weak lensing catalogue contains 723 background galaxies, corresponding to a density of $\sim$\,57 galaxies\,arcmin$^{-2}$, compared to other massive clusters such as $\sim$\,100\,arcmin$^{-2}$ from \citetalias{Jauzac2015_0416}, $\sim$\,42\,arcmin$^{-2}$ from \citetalias{Niemiec2023} in the deep \textit{Hubble Frontier Fields}, and $\sim$\,13\,arcmin$^{-2}$ from \citet{umetsu2012} in the Subaru sample.

\subsection{Grid mass modelling}
\label{sec:wl_modeling}
In contrast to the parametric strong-lensing analysis in the core of the cluster, in the outskirts, we use a non-parametric mass modelling technique to map the mass distribution
\citep{Jauzac2012,Jauzac2015_0416,Jauzac2016,Gonzalez2020,Niemiec2020,Niemiec2023}.

In MACS\,J1423, we here use a multiscale grid of potentials to decompose the matter distribution. The multi-scale grid method \citep{jullo2009,Jauzac2012} uses an irregular grid composed of Radial Basis Functions (RBFs) to add flexibility to the lensing mass reconstruction. RBFs are real value functions with radial symmetry fixed at all the nodes of the grid. The multi-scale grid is created from a smoothed lightmap of the cluster, and is recursively refined in the densest regions. The distribution of the cluster members is used to create a lightmap. We use \textsc{SExtractor} segmantation map to extract the light corresponding to these galaxies from the \textit{HST} F814W image and smooth it with a gaussian kernel to get a smoothed lightmap. On this smoothed lightmap, an hexagonal geometry of grid centred on the cluster core is adopted which recursively split into six equilateral triangles \citep[see][for more detail]{jullo2009}. Each RBF is modelled with a truncated isothermal mass distribution (TIMD, i.e., a circular version of dPIE used for our parametric mass modelling in Sect.\,\ref{sec:para_model}). This profile consists of a two component pseudo-isothermal mass distribution with a core radius, $s$, defined as the distance between an RBF and its closest neighbour, and a cut radius, $t$ \citep{jullo2009}. The true convergence field, $\kappa(\theta)$, is given by:
\begin{equation}
    \kappa(\theta) = \frac{1}{\Sigma_{\rm{crit}}}\Sigma_{i} v_{i}^{2} f \left(||\theta_{i} - \theta||, s_{i},t_{i}\right),
\end{equation}

where the RBF at the grid node, $\theta_{i}$, is defined as :

\begin{equation}
    f(R,s,t) = \frac{1}{2G}\frac{t}{t-s}\left(\frac{1}{\sqrt{s^{2} + R^{2}}} - \frac{1}{\sqrt{t^{2} + R^{2}}} \right).
\end{equation}

In a TIMD profile, the scaling factor, $v_{i}^{2}$, is the velocity dispersion at the centre of the gravitational potential, $i$. The cut radii of all grid potentials are fixed to $s=3\,t$ \citep{jullo2009}.

Following this, the shear field is written as

\begin{equation}
    \gamma_{1} (\theta) \, = \, \Sigma_{i} v_{i}^{2} \Gamma_{1}^{i}\left({||\theta_{i} - \theta_{j}}||,\, s_{i},\, t_{i} \right), 
\end{equation}

\begin{equation}
    \gamma_{2} (\theta) \, =\, \Sigma_{i} v_{i}^{2} \,\Gamma_{2}^{i}\left({||\theta_{i} - \theta_{j}}||,\, s_{i},\, t_{i} \right),
\end{equation}
where $\Gamma_{1}$ and $\Gamma_{2}$ are given in equation A8 in \citet{2007arXiv0710.5636E}. 

The ellipticity measurements in the weak lensing regime can be approximately given as a liner relation,
\begin{equation}
    \bm{e}  =  M_{\gamma v} \bm{v} +  \bm{n},    
\end{equation} 

where $\bm{v}$ contains the amplitude of the RBFs and $\bm{n}$ gives the intrinsic ellipticity and the noise in the shape measurements. The shear components of the matrix $\bm{M_{\gamma v}}$ containing the cross-contribution of each individual RBF to each individual weak lensing source are scaled by a ratio of angular diameter distances. The elements of this matrix for the two shear components are given by :
\begin{equation}
    \Delta_{1}^{(j,i)}\, = \,\frac{D_{LS,i}}{D_{OS,i}}\,\Gamma_{1}^{i}\left({||\theta_{i} - \theta_{j}}||,\, s_{i},\, t_{i} \right), 
\end{equation}
\begin{equation}
    \Delta_{2}^{(j,i)}\, =\, \frac{D_{LS,i}}{D_{OS,i}}\,\Gamma_{2}^{i}\left({||\theta_{i} - \theta_{j}}||,\, s_{i},\, t_{i} \right),
\end{equation}
where the elements corresponds to the contribution from each unweighted RBFs \textit{j} to the shear of image \textit{i}, and $\Gamma_{1}$ and $\Gamma_{2}$ are given in equation A8 in \citet{2007arXiv0710.5636E}..

\begin{figure}
    \includegraphics[width=\linewidth]{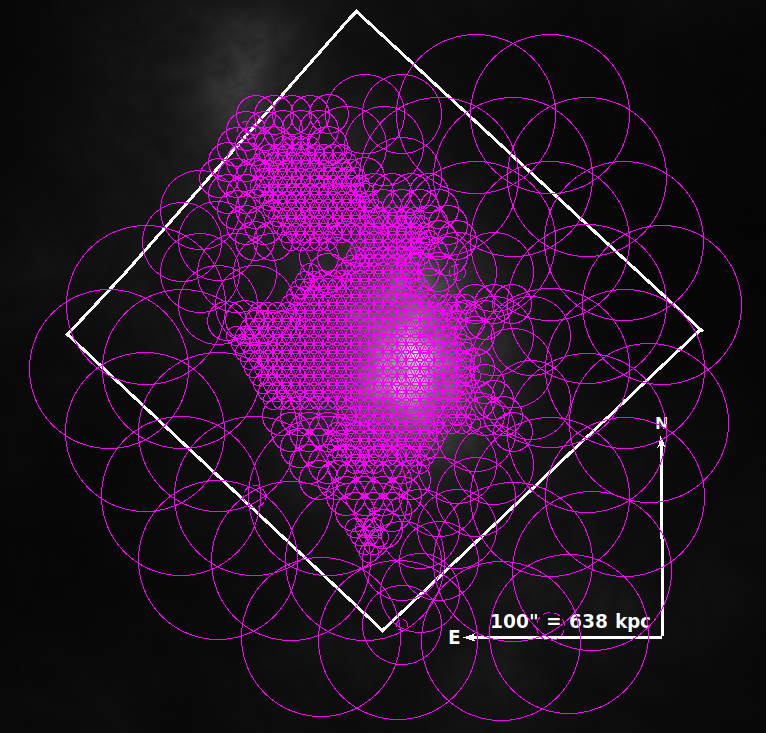}
    \caption{Smooth light map of MACS\,J1423. The non-parametric grid of potentials used for our weak-lensing analysis is highlighted with magenta circles. The grid is multi-scale, i.e. with a finer resolution in the densest regions of the cluster. The grid is composed of 162 RBFs, the densest regions modelled by RBFs with a core radius of 10.6\arcsec. The ACS field of view is shown by the white box.}
    \label{fig:grid}
\end{figure}

After testing different grid resolutions, we converge on an optimal grid composed of 628 RBFs. Core radii, \textit{s}, are ranging from 5\arcsec\ and 40\arcsec. The grid is constructed using the smoothed light map of the cluster which is made using publicly available scripts \footnote{\url{https://github.com/AnnaNiemiec/grid_lenstool}}. The resulting grid is shown in Fig.\,\ref{fig:grid} on the smoothed cluster light map. Each circle on the light map represents one RBF potential, and the densest region corresponds to \textit{s} = 5\arcsec. Our grid only model, covering the full ACS field (100$\arcsec$ or 640kpc), uses only background galaxies as constraints. The surface mass density for the corresponding model is shown in Fig.\,\ref{fig:sdp1}. The projected surface mass within the full ACS field is \textit{M}(${R}<640\,{\rm kpc})=(2.1\pm 0.8)\times$10$^{14}$\,M$_{\rm\odot}$.

\section{Combining Strong and Weak lensing}
\label{sec:sl+wl}
In order to obtain a full mass reconstruction of MACS\,J1423, we need to combine our parametric strong-lensing mass model with the non-parametric weak-lensing one. For that, we follow the sequential method described initially in \citetalias{Jauzac2015_0416} and then developed further in \cite{Niemiec2020} and \cite{Niemiec2023}.  First, the core is modelled using the strong lensing constraints as described in Sect.\,\ref{sec:sl}. To the best-fit parametric model of the core, the multi-scale grid of potentials described in Sect.\,\ref{sec:wl_modeling} is added in the outskirts of the cluster. 
The observed ellipticity from the sum of the two components of our model can be written as
\begin{equation}
    \bm{e_{m}}  =  M_{\gamma v}\bm{v} + \bm{e_{\rm{param}}} + \bm{n},
\end{equation} 
where \textbf{v} contains the amplitude of the RBFs of the multi-scale grid, \textbf{e$_{\rm m}$} = \textbf{(e$_{1}$,e$_{2}$)} corresponds to the individual shape measurements of the weak lensing sources, and \textbf{e$_{\rm param}$} is the fixed ellipticity from the best fit strong lensing model. \textbf{n} represents the galaxy shape noise component, and M$_{\gamma v}$ is the matrix which contains the contribution of individual RBFs. This matrix is presented in Sect.\,\ref{sec:wl_modeling}.

\begin{figure}
    \includegraphics[width=\linewidth]{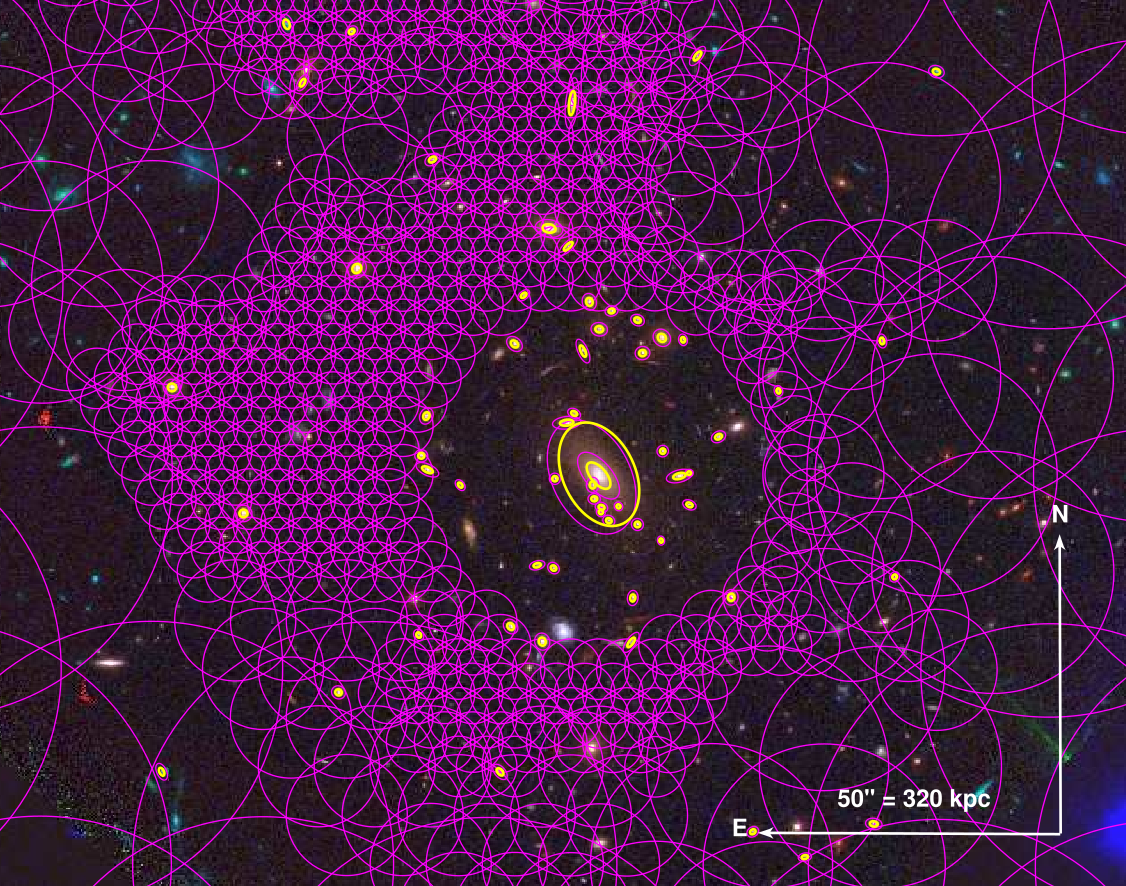}
    \caption{A zoom-in colour composite image on the core of MACS\,J1423 to highlight the combined strong and weak lensing analysis pursued in this work. The strong lensing mass components, fixed to the best fit parametric mass model, are shown in yellow, and the non-parametric grid potentials are shown in magenta.}
    \label{fig:combined_zoom}
\end{figure}

\begin{figure}
    \hspace{-0.5cm}
    \includegraphics[width=\linewidth]{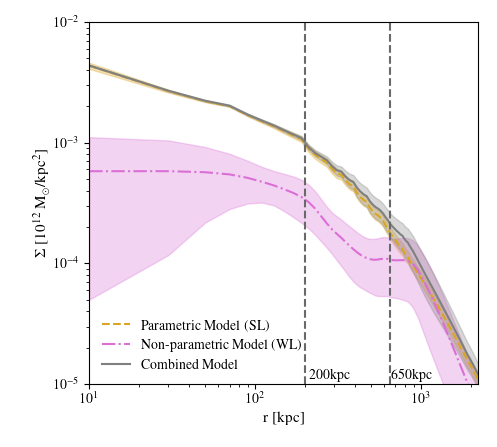}
    \caption{The surface mass density profiles of MACS\,J1423. The only strong lensing (sec\,\ref{sec:sl}) and weak lensing (sec\,\ref{sec:wl}) analysis are represented in yellow and magenta colour respectively. The profile in grey colour shows the combined model. The two vertical dark grey dashed lines represents the end of strong lensing and weak lensing regions respectively from left to right.}
    \label{fig:sdp1}
\end{figure}

As said before, we proceed to a sequential fit. The parametric model presented in Sect.\,\ref{sec:para_model} is fixed at its best-fit values.As shown in Fig.~\ref{fig:sdp1}, the error on the parametric model in the cluster core is much smaller than the error on the grid model. Consequently, fixing this component to the best-fit should not have a significant impact on the overall error propagation and model fitting process. The RBF amplitudes are then estimated from the weak lensing constraints. This also ensures that strong lensing constraints are not taken into account twice. Already optimised cluster scale dark matter halo and cluster member galaxies are combined with the RBF grid potentials. As explained in Sect.\,\ref{sec:wl_modeling}, grid potentials are removed from the cluster core due to the absence of weak lensing constraints in this region. A zoom-in combined parametric and non-parametric potentials is shown in Fig.\,\ref{fig:combined_zoom}. 
The projected surface mass density resulting from our combined strong and weak lensing mass model of MACS\,J1423 is shown in Fig.\,\ref{fig:sdp1}. We here also decompose the contribution of the strong and weak lensing optimisation. We measure a (combined strong and weak lensing) projected mass over the full \textit{HST}/ACS field of view ($R<100\arcsec$) of \textit{M}($R<640\,{\rm kpc})=(6.6\pm 0.6)\times 10^{14}$\,M$_{\rm \odot}$.

\begin{figure}
    \hspace{-0.5cm}
    \includegraphics[width=\linewidth]{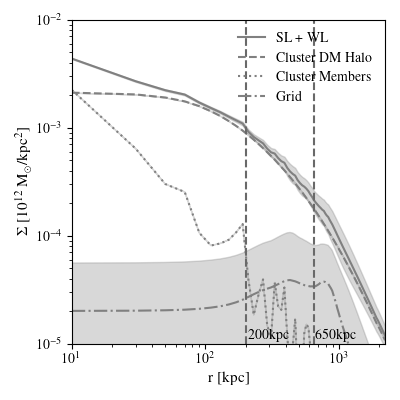}
    \caption{The surface mass density profiles of MACS\,J1423. The full combined strong and weak lensing model is represented by the solid line. The different components of this model i.e. cluster scale halo, cluster members and the grid potentials are shown in dashed, dotted and dashed-dotted lines respectively. The two vertical dark grey dashed lines represents the end of the strong and weak lensing regions respectively from left to right.}
    \label{fig:sdp2}
\end{figure}

\section{Discussion and Conclusion}
\label{sec:disc}
\subsection{Comparison between different models}
There are three main models presented in this paper : the parametric model for the cluster inner core, the non-parametric grid model the outskirts and the combined model accommodating both strong and weak lensing regions. Mass contours corresponding to all these three models are shown in Fig.\,\ref{fig:pot_contours}. In addition to these contours, we also show the smoothed X-ray surface brightness contours. A general trend in the elongation of the system is seen for all models even if some deviations can be seen in mass contours from the grid only model as compared to the parametric and combined models.  
Figure\,\ref{fig:sdp1} shows the 2-D surface density profile for our three models. As seen in Fig.\,\ref{fig:sdp1}, the amplitude of the grid only model is low compared to the parametric and combined model throughout the cluster region. In the core of the cluster, this is assignable to the absence of strong lensing constraints and the contribution from any cluster scale halo. However for the outer regions, either the grid only model is underestimating the mass or there is an overestimation from the parametric model. A plausible explanation to the later can be that outer regions of the cluster are affected by the elongation of the cluster scale halo (with $r_{\rm cut}$ = 1000\,kpc). To inspect this, we ran a parametric model including weak lensing constraints in the outskirts along with strong lensing constraints in the core. The presence of constraints (weak lensing) in the outer regions of the cluster can be used to constrain $r_{\rm cut}$ of the cluster scale halo. The difference in the surface density was not resolved even after constraining the cluster scale $r_{\rm cut}$. Similar problems are also seen in \citet{Niemiec2023} with a much more complicated cluster using same techniques as described in this paper. The reason for this difference in the amplitude between the grid only and combined model in the outskirts needs to be investigated. 

\begin{table*}
\caption{A summary table to compare the mass of MACS\,J1423 with previous works. The masses measured within 130\,kpc, 200\,kpc, 415\,kpc and 640\,kpc using different methods are quoted in units of $10^{14}$\,M$_{\rm\odot}$. The other works in the table corresponds to : \citet[][LR03]{2003ApJ...583..559L} and \citet[][SA07]{2007MNRAS.379..209S}, \citet[][Z11]{zitrin2011}, \citet[][L10]{Limousin2010}. We do not include our WL only model in the inner regions of the cluster i.e. for $R<130$\,kpc and $R<200$\,kpc due to lack of WL constraints in these regions. Similarly, we do not include our SL only model in outer regions of the cluster i.e. for $R<415$\,kpc and $R<640$\,kpc due to lack of SL constraints in these regions. }
\centering
\hspace*{-1cm} 
\begin{tabular}{c c c c c c c}
\hline\hline
 & Method & LR03 & SA07 & L10 & Z11 & This work \\
\hline\hline
\textit{R} < 130\,kpc & SL only & - & - & - & 1.3 $\pm$ 0.4 & 0.96 $\pm$ 0.01 \\
             & SL + WL & - & - & .. & - & 0.97 $\pm$ 0.02 \\
\hline
\textit{R} < 200\,kpc & SL only & - & - & .. & - & 1.82 $\pm$ 0.05  \\
             & SL + WL & - & - & .. & - &  1.85 $\pm$ 0.05 \\
\hline
\textit{R} < 415\,kpc & SZ & 5.0$_{-0.9}^{+3.1}$ & - & - & - & - \\
             & X-rays & - & 3.1$_{-0.7}^{+0.9}$ & - & - & - \\
             & WL only & - & - & - & - & 1.3 $\pm$ 0.4 \\
             & SL + WL only & - & - & 4.3$\pm$ 0.6 & - & 4.4 $\pm$ 0.3 \\
\hline
\textit{R} < 640\,kpc & WL only & - & - & - & - & 2.1 $\pm$ 0.8 \\
             & SL + WL only & - & - & ... & - & 6.6 $\pm$ 0.6 \\
\hline\hline
\end{tabular}
\label{tab:masses}
\end{table*}

\subsection{Comparison with previous works}

In \citetalias{Limousin2010}, mass modelling is slightly different than in this work, even if both analyses uses \textsc{Lenstool}. While we optimise the BCG independently from the other cluster galaxies (which are assume to follow a scaling relation), \citetalias{Limousin2010} assumed the BCG follows the scaling relation presented in equation\,\ref{eq:scaling_relation}. They measure a projected mass within 65\arcsec ($\sim$415\,kpc) of $M_{\rm L10}(R<415\,{\rm kpc})=(4.3\pm 0.6)\times 10^{14}$\,M$_{\rm\odot}$ which is within the errorbars of our strong + weak lensing analysis, \textit{M}($R< 415\,{\rm kpc})=(4.4\pm 0.3)\times 10^{14}$\,M$_{\rm\odot}$. 

\citetalias{zitrin2015} uses a different algorithm than in this work, LTM (Light-Traces-Mass), and its dPIE$+$NFW \citep[Navarro Frank White][]{nfw1996} version of it. \citetalias{zitrin2015} quotes a rms of 1.21\arcsec, and 1.47\arcsec\ for LTM and dPIE$+$NFW respectively, slightly higher than the rms of 0.9\arcsec obtained with our strong-lensing analysis. \citet{zitrin2011} measure a projected strong lensing mass enclosed within an effective Einstein radius of 20\arcsec\ ($\sim$130\,kpc) of $M_{Z11}(R<130\,{\rm kpc})=(1.30\pm 0.40)\times 10^{14}$\,M$_{\rm\odot}$. This is in good agreement with what we obtain $M(R<130\,{\rm kpc})=(0.96\pm 0.01)\times 10^{14}$\,M$_{\rm\odot}$ for our strong lensing model. 

The Sunyaev-Zel`dovich (SZ) effect analysis presented in \cite{2003ApJ...583..559L} quotes a mass of $M_{\rm LR03}(R<415\,{\rm kpc})=5.0_{-0.9}^{+3.1}\times 10^{14}$\,M$_{\rm\odot}$. The \textit{Chandra} X-ray analysis from \cite{2007MNRAS.379..209S} quotes a mass of $M_{\rm SA07}(R<415\,{\rm kpc})=3.1_{-0.7}^{+0.9}\times 10^{14}$\,M$_{\rm\odot}$. The SZ measurement fall within the error bars of mass estimates we obtain with this work. However, the mass estimate obtained from the X-ray analysis is lower than what we measure. \citetalias{Limousin2010} discuss this discrepancy and provide a possible explanation to the line of sight elongation of MACS\,J1423. A summary of all masses compared to in this work is given in Table\,\ref{tab:masses}.

\begin{figure}
    %\hspace{-0.5cm}
    \includegraphics[width=\linewidth]{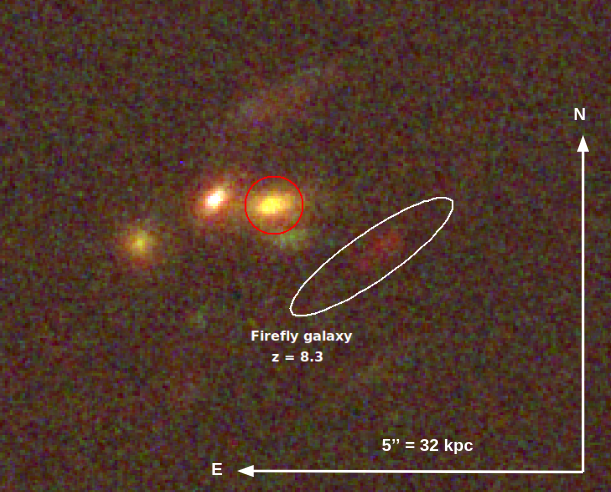}
    \caption{ The firefly galaxy at $z = 8.3 $ found by \citet{Mowla2024} is shown by the white ellipse. The galaxy highlighted in red may be giving a boost in the magnification locally. This \textit{HST} image is created using the F606W for blue, F814W for green and F160W for red colours.}
    \label{fig:firefly}
\end{figure}

Recently, \citet{Mowla2024} discovered a $z=8$ galaxy, calling it the Firefly Sparkle arc in MACS\,J1423 using JWST data. They measure magnification between 16-26, with the center of the arc being magnified by a factor of 24 using a strong lensing model. When compared to work presented in this paper, we measure a magnification of 6 with our strong lensing only model and a magnification of 17 with our strong + weak lensing model. The reason for this difference may be the fact that our models do not include a faint galaxy located near this lensed arc, which may be giving a boost in the magnification locally. Figure\,\ref{fig:firefly} shows the firefly galaxy in white and this local galaxy in red.

\begin{figure}
    \centering
    \hspace*{-1cm} 
    \includegraphics[width=\linewidth]{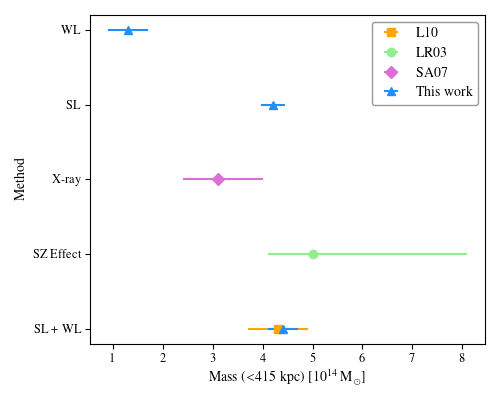}
    \label{fig:masses}
    \caption{A visual representation of the different masses given in table~\ref{tab:masses}. The methods used to estimate the enclosed mass within 415\,kpc is shown from previous studies : \citet[][L10]{Limousin2010}, \citet[][LR03]{2003ApJ...583..559L} and \citet[][SA07]{2007MNRAS.379..209S}.}
\end{figure}

\subsection{The baryonic mass distribution}
\subsubsection{Gas and Stellar mass distribution}
We follow the procedure presented in \cite{Beauchesne2024} to estimate the X-ray gas-density and gas-mass profiles for MACS\,1423 using \textit{Chandra} data (see section~\ref{sec:chandra}). The X-ray peak emission in the cluster is also measured using the publicly available code \textsc{pyproffit} \citep{Eckert2020}. We find that the X-rays and lensing distributions are aligned, i.e. the X-rays peak at the center of the cluster location of BCG. We measure a gas mass of $M_{\rm gas}(R<200\,{\rm kpc})=(0.16 \pm 0.0002)\times 10^{14}$\,M$_{\rm\odot}$. The gas mass is converted into gas fraction by taking the ratio between the gas mass and the mass of the cluster with 200\,kpc. We estimate a gas fraction of $f_{\rm gas}(R<200\,{\rm kpc}) = 0.08 \pm 0.01$. 
\\

To measure the stellar mass distribution, we follow the procedure described in \citet{Jauzac2015_0416}. We estimate the stellar masses of cluster galaxies using the relation log${(M_{\star}/L_{ \rm K}) = az + b}$, where \textit{z} is the redshift of the cluster, here $z=0.545$. The relation established by \cite{arnouts2007} for quiescent (red) galaxies in the VVDS survey \citep{LeFevre2005}, adopting a Salpeter initial mass function (IMF), and $L_{\rm K}$ is the luminosity in K-band. The parameters \textit{a} and \textit{b} are given by:
\begin{align*}
    \begin{split}
        a = -0.18 \pm 0.03, \\ 
        b = -0.05 \pm 0.03.
    \end{split}    
\end{align*}
We apply this relation to our 66 cluster member galaxies. First, we estimate the K-band luminosity of our galaxies observed in F814W band using the theoretical models from \cite{Bruzual2003}. Assuming a range of exponentially decaying star histories within the range $\tau$ = 0.1 - 2 Gyr, we estimate a typical colour $(m_{\rm F814W} - m_{\rm K}) = 1.5199$ (AB system). We measure a total stellar mass within the full ACS field of view, $M_{\star}(R<640\,{\rm kpc})=(5.7 \pm 0.5)\times 10^{12}$\,M$_{\rm\odot}$ and mass-to-light ratio of $M_{\star}/L_{\rm K} = 0.7 \pm 0.1 $\,M$_{\rm\odot}$/L$_{\rm\odot}$. 
The fraction of total mass in stars, $f_{\star}$, is estimated by taking the ratio between the total stellar mass and the total mass of the cluster derived from our combined strong and weak lensing analysis.
We find a stellar fraction of $f_{\star}$ = 0.006 $\pm$ 0.001 across the ACS field of view.

\subsubsection{Baryon fraction}
\begin{figure}
    %\hspace{-0.5cm}
    \includegraphics[width=\linewidth]{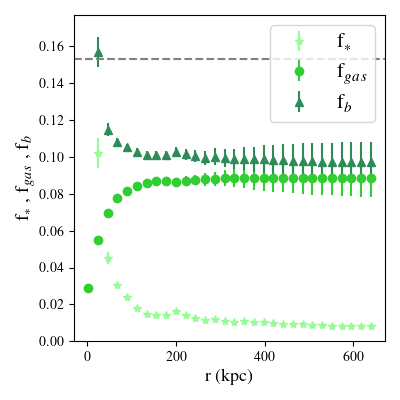}
    \caption{Fractions of stars, gas, and baryons within the ACS field of view, 640\,kpc from the center core. The grey dashed line represents the baryon fraction in cluster \citep{Krolewski2024}.}
    \label{fig:f_b}
\end{figure}

We use the derived gas and stellar masses to estimate the baryon fraction within the cluster. Figure~\ref{fig:f_b} shows the mass contribution from the gas and stars in the cluster as well as the baryon fraction. We see that the gas fraction increases as a function radius while the stellar fraction drops as we move away from the center. Within the ACS field of view, 640\,kpc from the center core, we measure a gas fraction of 0.09 $\pm$ 0.01 and a stellar fraction of 0.006 $\pm$ 0.001 The total baryon fraction within this radius is thus $f_{\rm b}$ = 0.095 $\pm$ 0.012. The cosmic baryon fraction measured by \citet{Planck2020}, is 0.154 and the baryon fraction in cluster is 0.153 \citep{Krolewski2024}.
The discrepancy might be due to several factors. Since the selection of cluster members mostly considered red galaxies and the ones spectroscopically confirmed are confined to the MUSE field (only 1\arcmin x 1\arcmin), the contribution from less massive star forming galaxies is not accounted for. Moreover, we are not including the stellar mass from the intra-cluster light (ICL) which can account for 10-40$\%$ of the total stellar mass in the cluster \citep{Gonzalez2007,Giodini2009,Lagana2013}.

\subsection{A possible test bench for dark matter}
Estimating the 3-D matter distribution in galaxy clusters plays an important role in testing the $\Lambda$CDM. Owing to the fact that lensing is sensitive to the integrated mass along the line of sight, the mass of a cluster is often overestimated due to the presence of mass concentrations not physically related to the cluster or to divergences from the spherical symmetry \citep{Gavazzi2005}. It was common practice to study dark matter distribution in clusters using X-rays and the assumption of spherical symmetry until departures from isothermality and spherical symmetry was found in \textit{XMM-Newton} and \textit{Chandra} observations in the core of several clusters. Moreover, the CDM paradigm predicts highly elongated axis ratios for dark matter halos in clusters \citep{wang2009}, contradicting the assumption of spherical shape. In a study by \citet{Gavazzi2005}, it was found that the discrepancy between the X-ray and gravitational lensing measurements can be explained by a non-spherical shape of clusters. Evidently, several studies suggest that prominent strong lensing observations are often accompanied by preferential elongations along the lines of sight or major axes preferentially oriented towards the lines of sight \citep{Peng2009,Corless2009}.  

The 3-D shape of the MACS\,J1423 has been investigated in the past by \citet{Morandi2012} and \citet{limousin2013}, concluding that the cluster is rather triaxial in shape and elongated along the line of sight. \citet{Morandi2012} find that the cluster is triaxial, with a dark matter halo axis ratio of 1.53$\pm$0.15 and 1.44$\pm$0.07, on the plane of sky and along the line of sight respectively. Their study suggested that the mass discrepancy between X-rays and lensing can be solved with such a geometry. As mentioned earlier, the mass estimate from our strong + weak lensing analysis is larger than the X-rays, which might be due presence of structure along the line of sight as noted by \citet{Limousin2010,Morandi2012} and \citet{limousin2013}.

The $\Lambda$CDM model predicts a cuspy central density profile for dark matter halos \citep{2012MNRAS.425.2169G,1996MNRAS.283L..72N}, while observations indicate a flattened core \citep{2010MNRAS.402...21N,2008ApJ...674..711S}. Understanding dark matter distribution is crucial for studying the formation of galaxies and clusters, and understanding the properties of dark matter. Precise mass measurements using gravitational lensing and ground-based spectra can help constrain central density profiles \citep{2011ApJ...728L..39N,Newman2013_I,Newman2013_II}. In particular, the presence of radial arcs in the core of clusters are a now well recognised and powerful tool to constrain the very central (${\rm R<}20$\,kpc) mass distribution in galaxy clusters \citep{1999PThPS.133....1H}. As discussed in Sect.\,\ref{sec:mul_img}, MACS\,J1423 has two radial arcs (images 1.5 and 4.5), making this cluster an excellent candidate to be used as a dark matter test bench following similar methods as the ones presented in \cite{2001ApJ...555..504W} and \citet{2013ApJ...765...24N}. The recent analysis by Cerny et al. 2024 (submitted to MNRAS) presents a new way to use radial arcs as dark matter probes, following the work from \citet{2013ApJ...765...24N}. Cerny et al. 2024 (submitted to MNRAS) utilize lower-quality \textit{HST} and VLT/MUSE KALEIDOSCOPE data, enabling systematic analysis without extensive high-resolution spectroscopy. Such analysis considering MACS\,J1423 as part of a cluster sample is being pursued by our team.

\subsection{Summary \& conclusion}
We present a combined strong and weak lensing mass model of the massive galaxy cluster MACS\,J1423 at $z=0.545$, one of 12 high-redshift MACS clusters \citep[][]{2001ApJ...553..668E}. 
We model the cluster using a sequential fit which is composed of two different steps: \textit{(i)} the modelling of the strong-lensing region using a parametric mass model for which we assume that light traces mass, and \textit{(ii)} the modelling of the weak-lensing region using a multi-scale grid of potentials in order to include more flexibility in the modelling of this region where the lensing signal is much more diffuse than in the core.

Our strong lensing model is optimised using four multiple image systems, of which system 3 is spectroscopically confirmed with MUSE. We take advantage of VLT/MUSE observations of the core of MACS\,J1423, taken as part of the KALEIDOSCOPE survey which allowed us to confirm existing spectroscopic measurements of multiple image systems together with obtaining new spectroscopic measurements for 39 cluster galaxies. We compile a cluster galaxy catalogue composed of 66 galaxies using the red sequence technique and calibrated with our VLT/MUSE spectroscopic redshift measurements. Our best-fit mass model recovers the multiple image positions with a rms of 0.90\arcsec, and yields a projected enclosed mass of \textit{M}($R<200\,{\rm kpc})=(1.82\pm 0.05)\times 10^{14}$\,M$_{\rm\odot}$. 

In contrast to the parametric model in the core, since the positions of mass clumps are not known in the outskirts, a more flexible, non-parametric grid of mass potentials is used. The weakly lensed background galaxies, carrying the weak lensing signal, are used as constraints in the outskirts. We use 723 background galaxies selected within the \textit{HST}/ACS field of view, i.e. a density of 57\,gal.arcmin$^{-2}$. The weak lensing only model yields a projected enclosed mass of \textit{M}($R<640\,{\rm kpc})=(3.8\pm 0.8)\times$10$^{14}$\,M$_{\rm\odot}$.

To combine strong and weak lensing analyses, we use the sequential fit method as described in \cite{2015MNRAS.452.1437J} and \cite{Niemiec2020,Niemiec2023}. To the parametric best-fit model of the cluster core, a non-parametric grid is added to model outskirts of clusters. The combined strong and weak lensing analysis of MACS\,J1423 yields a mass of \textit{M}($R<200\,{\rm kpc})=(1.85\pm 0.05)\times 10^{14}$\,M$_{\rm\odot}$ in the core, and an enclosed mass of \textit{M}($R<640\,{\rm kpc})=(6.6\pm 0.6)\times 10^{14}$\,M$_{\rm\odot}$ within the full \textit{HST}/ACS field of view ($R<$100\arcsec).

The high mass, strong lensing power and relaxed dynamical state of MACS\,J1423 are powerful criteria to use as a possible dark matter test bench. The mass discrepancy between our lensing analysis and X-rays suggests elongation along the line of sight, indicating triaxial geometry of the cluster. This can be used to further investigate and test the $\Lambda$CDM paradigm of the Universe. Furthermore, the high redshift of the cluster, helps in understanding of the formation and evolution of structure in the early Universe.

\section*{Acknowledgements}
NP is supported by the Science and Technology Facilities Council (STFC, grant number ST/S505365/1, ST/V506643/1 and ST/W507428/1). 
MJ is supported by the United Kingdom Research and Innovation (UKRI) Future Leaders Fellowship `Using Cosmic Beasts to uncover the Nature of Dark Matter' (grant number MR/S017216/1 and MR/X006069/1). NP, AN, DL and GM acknowledge support from the UKRI FLF grant number MR/S017216/1 and MR/X006069/1 as well. ML acknowledges the Centre National de la Recherche Scientifique (CNRS) and the Centre National des Etudes Spatiale (CNES) for support. This reserach was supported by the \textit{International Space Science Institute} (ISSI) in Bern, through the ISSI International Team project \#476 ("Cluster Physics From Space To Reveal Dark Matter"). 

%%%%%%%%%%%%%%%%%%%%%%%%%%%%%%%%%%%%%%%%%%%%%%%%%%
\section*{Data Availability}
The CLASH \textit{HST} data used for this work are available on the MAST archive in the CLASH repository for MACS\,J1423\footnote{https://archive.stsci.edu/missions/hlsp/clash/macs1423/}. The lensing mass models and other data products will be shared by authors on request.

\bibliographystyle{mnras}
\bibliography{ref.bib} % if your bibtex file is called example.bib

% Don't change these lines
%\bsp	% typesetting comment
\label{lastpage}
\end{document}